\documentclass[intlimits,twoside,a4paper]{article}

\usepackage{amsmath,amssymb}
\usepackage{graphicx}

\usepackage[T2A]{fontenc}
\usepackage[cp1251]{inputenc}

\usepackage[eqsecnum]{cmpj2}

\issue{2014}{17}{3}{33702}
\doinumber{10.5488/CMP.17.33702}

\title[Exciton condensation]%
{Exciton condensation in quantum wells. Exciton hydrodynamics. The effect of localized states%
}

\author[V.I. Sugakov]{V.I. Sugakov}
\address{
Institute for Nuclear Research, 47 Nauky ave., 03680 Kyiv, Ukraine
}

\date{Received April 29, 2014}
\authorcopyright{V.I. Sugakov, 2014}

\begin{document}

\maketitle

\begin{abstract}
The hydrodynamic equations for indirect excitons in the double
quantum wells are studied taking into account 1) a possibility of
an exciton condensed phase formation, 2) the presence of
pumping, 3) finite value of the exciton lifetime, 4)
exciton scattering by defects. The threshold pumping emergence of
the periodical exciton density distribution is found.
The role of localized and free exciton states is analyzed in the
formation of emission spectra.

\end{abstract}

\section{Introduction}

Phase transitions in systems of unstable particles are  specific examples of non-equilibrium phase
transitions and processes of self-organization \cite{Hak83}.
 In such a system, particles are created by external sources and disappear due to different reasons. If there is an attractive interaction
  between the particles, they may create a condensed phase during their lifetime.  A steady state may arise in a system if the number
   of the created particles  in the unit time is equal to  the number of the disappeared particles. This state is stationary, but it is not equilibrium. The
following examples of such  systems with unstable particles may be
presented: 1) dielectric exciton liquid in crystals; 2)
electron-hole liquid in semiconductors; 3) highly excited gas with
many excited molecules; 4) vacancies and interstitials in a cryatal
created by nuclear irradiation; 5) quark glyuon plasma and
others. The finite value of the particles determine some
peculiarities of phase transitions in such systems. The main
peculiarities are as follows: a) a phase transition in a system
of unstable particles may occur if the lifetime is larger than
some critical value; b) in the presence of parameters at which two phases
coexistent, the sizes of the regions of condensed phases of unstable
particles are restricted: c) there is strong spatial correlation
between different regions of condensed phases, that is why
periodical structures may arise.

 The present paper is devoted to an investigation of self-organization processes of the exciton system in semiconductor quantum
 wells. The appearance of  periodical dissipative structures in exciton systems at
 the irradiations greater than some critical value was shown in the
 work \cite{Sug86,Sug98}. Experimental observation of a periodical
 distribution of the exciton density was obtained in \cite{But02, Gor06} in a system of
 indirect excitons in semiconductor double quantum wells. An
 indirect exciton consists of an electron and hole separated over two wells by an electric field.
 Due to the damping of the electron-hole recombination, indirect excitons  have  a large
 lifetime which makes it possible to create a high  concentration of excitons  at small pumping and
  to study the manifestation of the effects of exciton-exciton
  interaction. The authors of the paper  \cite{But02} observed the emission from a double quantum
  well on the basis of AlGaAs system in the form of periodically situated islands
  along  the ring around  the laser spot. In the paper \cite{Gor06}, in which the
excitation of a quantum well was carried out through a window in
a metallic electrode,  a periodical structure of the islands situated along the
ring under the perimeter of the window was found  in the luminescence
spectrum. The islands were observed at a
frequency that corresponds to the narrow line arising at
a threshold value of pumping \cite{Lar00}.  Afterwards, spatial
structures of exciton density distributions were observed in
a single wide quantum well \cite{Tim06}, in different types
of electrodes that create a periodical potential \cite{Rem09}
or have windows in the shape of a rectangle, two circles
and others \cite{Dem11,Gor12}.

The phenomenon of a symmetry loss and the  creation of
structures in the emission spectra of indirect excitons urged
a series of theoretical investigations
\cite{Lev05,Par07,Sap08,Liu06,Muk10,Wil12,And13, Bab13}. The
main efforts were directed towards the verification of a fundamental
possibility of the appearance of the periodicity of the exciton
density distribution. A specific explanation of the experiment is
presented in two works \cite{Lev05,Wil12} with respect to only one
experiment \cite{But02}.The authors of the work \cite{Lev05}
considered the instability that arises under the kinetics of  level
occupations by the particles with the Bose-Einstein statistics.
Namely, the growth of the occupation of the level with zero moment
should stimulate the transitions of excitons to this level. However,
the density of excitons was found greater, and the temperature was
found lower than these values observed in the experiments. In the
paper \cite{Wil12} the authors did not take into  account the
screening between the charges at macroscopic distances.

  Our model is based on the appearance of self-organization processes in non-equilibrium systems for  excitons with attractive
 interaction shown in \cite{Sug86,Sug98}.
     Investigations  performed in this model \cite{Sug05,Sug06,Cher06,Sug07,Cher07,Cher09,Cher12}  gave the
     explanations of spatial structures and their temperature
     and pumping dependencies obtained in
  different experiments \cite{But02,Gor06,Rem09,Dem11,Gor12}.
  Theoretical approaches of the  works \cite{Sug05,Sug06,Cher06,Sug07,Cher07,Cher09,Cher12}
are based on the following assumptions.

\begin{enumerate}
\item There is an exciton condensed phase caused by the attractive
interaction  between excitons.  The existence  of attractive
interaction between excitons is confirmed by the calculations of
biexcitons  \cite{Tan05,Schin08,Mey08,Lee09}, and  by
investigations of a many-exciton system \cite{Loz96}. Nevertheless,
there is an experimental work \cite{Yan07}, where the authors
 explain their experimental results by the existence of
a repulsion interaction between excitons. These results come into
conflict with our suggestion regarding the attractive interaction.  We
shall remove this contradiction in section~3.

\item  The finite value of the exciton lifetime plays an important
role in the formation of a spatial distribution of exciton
condensed phases.  As usual, the exciton lifetime
significantly exceeds the duration  of the establishment of a local
equilibrium.  However, it is necessary to take into account the finiteness of the
exciton lifetime  in the study of spatial
distribution phases in two-phase systems, because the exciton
lifetime is less than the time of the establishment of
equilibrium between phases. The latter is determined by slow
diffusion processes and  is long.

\end{enumerate}

Two approaches of the theory of
phase transitions were used while  developing the theory: the model of nucleation
(Lifshits-Slyozov) and the model of spinodal decomposition (Cahn-Hillart). These models were generalized to the particles with
the finite  lifetime, which is important
 for interpretation of the experimental results.  The involvement of Bose-Einstein
 condensation for excitons was
  not required in order to explain the experiments.

 In the present paper, the hydrodynamic equation for excitons is
investigated for the case of excitons being in a condensed phase. The
appearance of an instability of the uniform distribution of the
exciton density and the development of nonhomogeneous structures
are studied. The effect of defects on spectral positions
of the emission spectra of both gas and condensed phases is
analysed as well.

\section{Analysis of hydrodynamic equations of exciton condensed phase}

Hydrodynamic equations of excitons were obtained and analysed
in the work \cite{Lin92}. Hydrodynamic  equations of excitons
 generalizing the Navier-Stokes equations that take into account the finite exciton lifetime, the pumping
  of exciton, the  existence of an exciton condensed phase and the presence of defects were developed in \cite{SugUJP}.
  In the paper, we make some analysis of these equations.

  The system is described by the exciton density   $n \equiv n(\vec {r},t)$ and by the velocity of the exciton liquid $\vec
{u} \equiv \vec {u}(\vec {r},t)$. The
  equations for conservation of the exciton density and for the movement of
  exciton  density are basic for the exciton hydrodynamic
  equations.
\begin{eqnarray}
\label{eq8}
&&\frac{\partial n}{\partial t} + \textrm{div}(n\vec {u}) = G -
\frac{n}{\tau _\textrm{ex}}\,,\\
\label{eq9}
&&\frac{\partial mnu_i }{\partial t} = - \frac{\partial
\Pi _{ik} }{\partial x_k } - \frac{mnu_i }{\tau _\textrm{sc} }\,,
\end{eqnarray}
where $G$ is the pumping (the number of excitons created for unit
 time in unit area of the quantum well), $\tau_\textrm{ex}$ is the
 exciton lifetime,$m$ is the exciton mass, $\Pi _{ik} $ is the
tensor of density of the exciton flux
\begin{equation}
\label{eq10} \Pi _{ik} = P_{ik} + mnu_i u_k - {\sigma }'_{ik} \,,
\end{equation}
\noindent where
 $P_{ik}$  is the pressure tensor,
${\sigma }'_{ik}$ is the viscosity tensor of tension. In the
equation (\ref{eq9}), we neglected the small momentum change
caused by the creation and the annihilation of excitons.

 Introducing coefficients of viscosity  and using
(\ref{eq8}), equation (\ref{eq9}) may be rewritten in the form
\begin{eqnarray}
\label{eq11} \rho \left[ {\frac{\partial u_i }{\partial t} +
\left( {u_k \frac{\partial }{\partial x_k }} \right)u_i } \right]
= - \frac{\partial P_{ik} }{\partial x_k } + \eta \Delta u_i
+ (\varsigma + \eta / 3)\left( {\frac{\partial
}{\partial x_i }} \right)\textrm{div}\vec {u} - \frac{\rho u_i }{\tau _\textrm{sc}
}\,,
\end{eqnarray}
where $\rho=mn$  is the mass of excitons in the unit volume.

  We assume that the state of the local
equilibrium is realized and the state of the system may be
described by free energy, which depends on a spatial coordinate.
Let us present the functional of the free energy in the form
\begin{equation}
\label{eq12} F = \int {\rd\vec {r}\left[ {\frac{K}{2}\left(\vec{\nabla} n\right)^2 +
f(n)} \right]} .
\end{equation}

 At the given presentation of free energy, the pressure
tensor is determined by the formula \cite{Swi96}
\begin{equation}
\label{eq13} P_{\alpha \beta } = \left[ {p - \frac{K}{2}\left(\vec
{\nabla }n\right)^2 - Kn\Delta n} \right]\delta _{\alpha \beta } +
K\frac{\partial n}{\partial x_\alpha }\frac{\partial n}{\partial
x_\beta }\,,
\end{equation}
\noindent where $p = n{f}'(n) - f(n)$ is the equation of the
 state, $p$ is the isotropic pressure.

Taking into account  (\ref{eq13}) and introducing coefficients of
viscosity, we finally  rewrite the equation (\ref{eq9}) in the
form
\begin{eqnarray}
\label{eq14} \frac{\partial u_i }{\partial t} + u_k \frac{\partial
u_i }{\partial x_k } + \frac{1}{m}\frac{\partial }{\partial x_i
}\left( { - K\Delta n + \frac{\partial f}{\partial n}} \right) +
\nu \Delta u_i
+ (\varsigma / m + \nu / 3)\left(
{\frac{\partial }{\partial x_i }} \right)\textrm{div}\vec {u} + \frac{u_i
}{\tau _\textrm{sc} } = 0.
\end{eqnarray}
Equations (\ref{eq8}), (\ref{eq14}) are the equations of
the hydrodynamics for an exciton system.  It follows from the
estimations, made in the work \cite{Lin92},
 that the terms with the viscosity coefficients are small and we shall neglect them.

 In the case of a steady state irradiation, the
equations (\ref{eq8})  and (\ref{eq14}) have the solution $n=G\tau$, $u=0$.
  To study the stability of the uniform solution we consider, that the
 behavior of a small fluctuation of the exciton density and the
 velocity from these values are as follows: $n\rightarrow n+\delta n\exp[\ri\vec{k}\cdot\vec{r}+\lambda(\vec{k})
 t]$, $u=\delta u\exp[\ri\vec{k}\cdot\vec{r}+\lambda(\vec{k})
 t]$.  Having substituted  these expressions in equations (\ref{eq8}),  (\ref{eq14}), we obtain, in the linear approximation with respect to
 fluctuations, the following expression
\begin{eqnarray} \label{eq14aa} \lambda_{\pm}(\vec{k})=\frac{1}{2}\left[-\left(\frac{1}{\tau_\textrm{sc}}+\frac{1}{\tau_\textrm{ex}}\right)
\pm\sqrt{\left(\frac{1}{\tau_\textrm{sc}}-\frac{1}{\tau_\textrm{ex}}\right)^2-
\frac{4k^2n}{m}\left(k^2K+\frac{\partial^2 f}{\partial n^2}\right)} \right].
\end{eqnarray}

It follows from (\ref{eq14aa}) that both parameters
$\lambda_{\pm}(\vec{k})$ have  a negative real part at small and
large values of vector $\vec{k}$ and, therefore, the uniform
solution of the hydrodynamic equation is stable.  The instability
with respect to a formation of nonhomogeneous structures arises at
some threshold value of exciton density and at some  critical
value of the wave vector, when ${\partial^2 f}/{\partial n^2}$
becomes negative. The analysis of the equation (\ref{eq14aa}) gives
the following expression for the critical values of the wave
vector $k_\textrm{c}$ and the exciton density $n_\textrm{c}$
\begin{eqnarray} \label{eq14ab}
&&k_\textrm{c}^4=\frac{m}{Kn_\textrm{c}\tau_\textrm{sc}\tau_\textrm{ex}}\,,\\
\label{eq14ac}
&&\frac{k_\textrm{c}^2n_\textrm{c}}{m}\left(k_\textrm{c}^2K+\frac{\partial^2 f(n_\textrm{c})}{\partial
n_\textrm{c}^2}\right)+\frac{1}{\tau_\textrm{sc}\tau_\textrm{ex}}=0.
\end{eqnarray}

For stable particles ($\tau_\textrm{ex}\rightarrow\infty$), the equations
(\ref{eq14ab}), (\ref{eq14ac}) give the condition ${\partial^2
f}/{\partial n^2}=0$, which is the condition for spinodal decomposition
for a system in the equilibrium case.

Depending on parameters, the equations (\ref{eq8}), (\ref{eq14}) describe
the ballistic  and diffusion movement of the exciton system.  The
relaxation time $\tau _\textrm{sc}$ plays an important role in the
formation of the exciton movement.  Due to the appearance of
nonhomogeneous structures, there exist exciton currents  in a system
 ($\vec{j}=n\vec{u}\neq 0$) even under
 the uniform steady-state pumping. Excitons are moving from the regions having a small exciton density
 to the regions having a high density. In the present paper, we shall consider
the spatial distribution of exciton density and exciton current in
the double quantum well under steady-state pumping.   In this case,
the exciton carrent is small and we assume the existence of the
following conditions
\begin{align}
\label{eq14a}
\frac{\partial u_i }{\partial t}
&\ll u_i/\tau _\textrm{sc}\,,\\
\label{eq14b}
u_k \frac{\partial u_i }{\partial x_k } &\ll u_i /\tau
_\textrm{sc}\,.
\end{align}
Particularly, the equation (\ref{eq14a}) holds in the study of the
steady-state exciton distribution. The fulfilment of
equation (\ref{eq14b}) will be shown later following some numerical
calculations.

Using  the conditions (\ref{eq14a}) and (\ref{eq14b}),
we obtain from equation (\ref{eq14}) the value of the
velocity  $\vec {u}$
 \begin{equation}
\label{eq15} \vec {u} = - \frac{\tau _\textrm{sc}}{m}\vec {\nabla }\left(
{ - K\Delta n + \frac{\partial f}{\partial n}} \right).
\end{equation}
  As a result,  the equation for the exciton density current  may be presented in the form
\begin{equation}
\label{eq16} \vec {j} = n\vec {u}=- M\nabla \mu,
\end{equation}
\noindent  where $\mu = \delta F/\delta n$ is the chemical
potential of the system,
 $M = nD/\kappa T$ is the mobility, $D =
\kappa T\tau _\textrm{sc} /m$ is the diffusion coefficient of excitons.

Therefore, the equation for the exciton density (\ref{eq8}) equals
\begin{eqnarray}
\label{eq17} \frac{\partial n}{\partial t} = \frac{D}{\kappa T}
\left( -
Kn\Delta ^2n - K\vec {\nabla }n \cdot \vec {\nabla }\Delta n\right)
 +\frac{D}{\kappa T}\vec {\nabla }\cdot \left(
n\frac{\partial ^2f}{\partial n^2}\vec {\nabla }n \right)  + G -
\frac{n}{\tau _\textrm{ex} }\,.
\end{eqnarray}

Just in the form of (\ref{eq17}),  we investigated a spatial
distribution of the exciton density at exciton condensation using
the spinodal decomposition approximation by choosing different
dependencies $f$ on $n$ \cite{Cher06, Cher07, Cher12}. Thus, our
previous consideration of the problem corresponds to the diffusion
movement of hydrodynamic equations (\ref{eq8}), (\ref{eq14}).
For the  system  under study, a condensed phase appears if  the function $f(n)$ describes a phase transition.
In the papers mentioned above, the examples of such dependencies were given. Here, we  analyse another dependence $f(n)$, which is also often
used in the theory of phase transitions
\begin{equation}
\label{eq18} f = \kappa Tn\left[\ln (n / n_a ) - 1\right] + a\frac{n^2}{2} +
b\frac{n^4}{4} + c\frac{n^6}{6}\,,
\end{equation}
where $a$, $b$, $c$ are constant values.  Three last
terms in the formula (\ref{eq18}) are the main terms. They arise
due to an exciton-exciton interaction and describe the phase
transition.
 The first term was introduced in order to describe the system
 in a space, where the exciton concentration
is small (this is important if such a region exists in a system).
At an increase of the exciton density, the term $a{n^2}/{2}$
manifests itself firstly. It contributes the $an$ value
to the chemical potential. In our
system, the origin of this term is connected  with the dipole-dipole interaction, which should become
apparent at the beginning with the growth of the density due to
its long-range nature. To estimate $a$  for the
dipole-dipole exciton interaction in double quantum well we may use the plate
capacitor formula $an=4\pi e^2 dn/\epsilon$, where $d$ is the
distance between the wells, $\epsilon$ is the dielectric constant.
This formula is usually used  to determine the exciton
density from the experimental meaning of the blue shift of the
frequency of the exciton emission with the rise of the density. It
follows from the formula that $a=4\pi e^2 d/\epsilon$.  When the
exciton density grows, the last two terms in (\ref{eq16}) begin
to play a role. The existence of a condensed phase requires that
the value $b$ should be negative ($b<0$). For stability of a system, at
large $n$, the parameter $c$ should be positive ($c>0$). It is
assumed in the model that the condensed phase arises due to the
exchange and Van der Waals interactions.  The calculations show
that in some region of distances between the wells,  these
interactions exceed the dipole-dipole repulsion.

 Let us introduce  dimensionless parameters:
$\tilde {n} = n / n_0 ,$ where $n_0 = \left( {a / c} \right)^{1 /
4}$, $\tilde {b} = b / (ac)^{1 / 2}$, $\tilde {\vec {r}} = \vec{r}
/ \xi $, where $\xi = \left( {K / a} \right)^{1 / 2}$ is the
coherence length, $\tilde {t} = t / t_0 $, where $t_0 =
{\kappa TK}/({Dn_0 a^2})$, $D_1 = {\kappa T}/({an_0 })$,
$\tilde {G} = Gt_0/n_0 $, $\tilde {\tau }_\textrm{ex} = \tau / t_0 $. As
a result, the equation (\ref{eq17}) is reduced to the form
(hereinafter the symbol $\sim $ will be omitted in the equation)
\begin{eqnarray}
\label{eq19} \frac{\partial n}{\partial t} &=& D_1 \Delta n -
n\Delta ^2n + n\Delta n\left(1 + 3bn^2 + 5n^4\right)
\nonumber \\
&&- \vec
{\nabla }n\cdot\vec {\nabla }\Delta n+\left(\vec {\nabla }n\right)^2\left(1 +
9bn^2 + 25n^4\right) + G - \frac{n}{\tau _\textrm{ex}}\,.
\end{eqnarray}

\begin{figure}[!t]
\centerline{
\includegraphics[width=8.6cm]{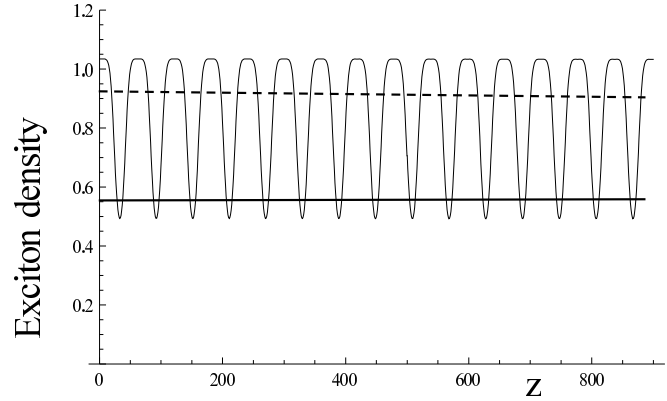}
}
\caption{
The spatial dependence of the exciton density at a different value of the pumping:
for a continues line   $G=0.0055$, for a periodical line $G=0.008$, for a dashed line
$G=0.0092$. $D_1=0.03$, $b=-1.9$.}
\label{fig1}
\end{figure}

The  solutions of the equation  (\ref{eq19})
are presented in figure~\ref{fig1} for the one-dimensional case [$n(\vec {r},t) \equiv n(z,t)$] for three values of the steady-state uniform pumping.

\begin{figure}[!t]
\centerline{
\includegraphics[width=8.6cm]{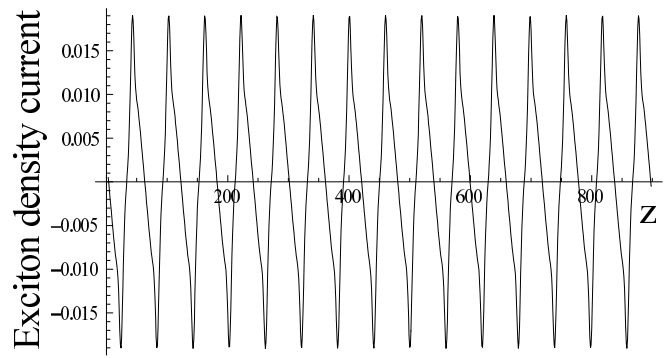}
}
\caption{
 The spatial dependence of the exciton current at  $G=0.008$,  $D_1=0.03$, $b=-1.9$.   }
\label{fig2}
 \end{figure}

The solutions are obtained  at the initial conditions  $n(z,0) =
0$ and the boundary conditions $n'(0,t) = n'(L,t) = n''(0,t) =
n''(L,t) = 0$, where $L$ is the size of a system.   The
periodical solution exists in some interval of the pumping
$G_{\textrm{c}1} < G < G_{\textrm{c}2} $.   At specified parameters, the periodical
solution exists at $0.0055 < G < 0.0092$. Outside this region, the
solution describes a uniform system: the gas phase at a low
pumping and the condensed phase at a large pumping. The upper part of the periodical distribution corresponds to a condensed phase,
  the lower part corresponds to the gas phase.
 The size of the condensed phase increases with the change of the pumping from  $G_{\textrm{c}1}$ to $G_{\textrm{c}2}$.
  Figure~\ref{fig2} shows the spatial dependence of the exciton current
 calculated by the formula (\ref{eq16}). The current equals zero in
 the centers of the condensed and gas phases and it has a maximum in the
 region of a transition from the condensed phase to the gas
 phase.
Let us do some estimations.  The results for the
currents in figure~\ref{fig2} are presented in dimensionless units:
$\tilde{j}=j/j_0$, where $j_0=n_0u_0$,
$u_0=(\tau_\textrm{sc}n_0a)/(m\xi)$ is the unit of the velocity. The exciton density
is presented in figure~\ref{fig1} in dimensionless units ($\tilde{n}=n/n_0$). It is seen in figure~\ref{fig1} that $\tilde{n}\sim 1$,
and the magnitude of $n$ is of an order of $n_0$. Thus, for
estimations we may assume that $n_0a$ corresponds to  the shift
of the luminescence line with an increase of the exciton density,
the magnitude of $\xi$ is of the order of the size of the
condensed phase. For the following magnitudes of parameters $\tau_\textrm{sc}=10^{-11}$~{s}, $n_0a=2\cdot10^{-3}$~{eV}, $m=2\cdot10^{-28}$~{g},
$\xi=2\cdot 10^{-4}$~{cm}, we obtain $u_0\sim 10^6$~{cm/s}. According
to calculations (see figure~\ref{fig2}), the magnitudes of the current and
the velocity are   two orders of magnitude less than their
units $j_0$ and $u_0$, so the condition  $u\sim10^{4}$~{cm/s} takes place.   In order to verify the fulfilment of the condition (\ref{eq14b}),
let us suppose that
($\partial u_i)/(\partial x_k\sim u/l$), where $l$ is
the period of a structure. It  follows from experiments \cite{But02,Gor06} that $l\sim (5\div10)$~\textmu{}m. Using these data we see
that the condition ($\ref{eq14b}$) is very well satisfied. This condition is violated  at $\tau_\textrm{sc}\geqslant 10^{-9}$~{s}.
Therefore, the formation
of nonuniform exciton dissipative structures  in a double
quantum well occurs due to the diffusion movement of excitons.
To prove the main hydrodynamic equation (\ref{eq17}), the
last term in equation (\ref{eq9})is of importance. It describes the
loss of the momentum due to the scattering of excitons by defects
and phonons. It is this term that describes the processes that cause a
decay of the exciton flux. From the viewpoint  of a possibility
of the appearance of superfluidity, the situation for excitons is
more complicated than that for  the liquid helium and for the
atoms of alkali metals at ultralow temperatures.
 In the latter systems, the phonons (movement of particles) are an intrinsic
 compound part of the system spectrum, the interaction between
 phonons (particles) is the interaction between the atoms of a system and does not cause the change
 of the complete momentum of a system and  its movement as a whole. Phonons and defects
 for excitons are  external subsystems that brake the exciton movement. Therefore,
  to  create the exciton superfluidity, it is needed that the value of $\tau _\textrm{sc}$
 should grow significantly. This is possible for exciton polaritons
that weakly interact with phonons; moreover, there is a certain
experimental evidence on an observation of the polariton
condensation \cite{Kas06}. For indirect excitons, the critical
temperature of a superfluid transition is strongly lowered by
inhomogeneities \cite{Ber06,Bez11}. Thus, the question regarding the
possibility of the superfluidity existence for indirect
excitons on the basis of  AlGaAs system is open.

Thus, the peculiarities observed at large densities of indirect excitons
may be explained by phase transitions in a system of particles having attractive
interactions and by the finite value of the lifetime without an involvement of the Bose-Einstein condensation.

\section{Distribution of excitons between localized and delocalized
states}

According to the experimental results \cite{Yan07},  the
frequency of the emission from the islands of a condensed phase,
where the exciton density is large, is higher than the frequency
of emission from the region between islands, in which the density
is less. The authors made the conclusion \cite{Yan07} that
the interaction between excitons is repulsive, and, therefore, the
formation of a condensed phase by attractive interaction is
impossible.  This contradicts the main assumption of our works
\cite{Sug05,Sug06,Cher06,Sug07,Cher07,Cher09,Cher12}, though these
works explain many experiments. Now, we remove this contradiction
taking into account the presence of localized excitons.

 The localized states arise due to the presence of residual donors, acceptors, defects,
and inhomogeneous thickness of the wells.  Their existence is
confirmed by the presence of an emission at the frequencies less than
the frequency of the exciton band emission and by broadening of
exciton lines. At a low temperature and at a small pumping, the
main part of the emission band consists of the emission from
defect centers, while the part of the exciton emission grows with
an increased pumping.  Now, we consider the relation between the
contribution to the emission band intensity from free excitons and from
the excitons (pairs of electrons and holes) localized on the defects.
 We assume that the localized states are saturable, namely, every center may
capture a restricted number of electron-hole pairs. In our
calculations we assume  that only a single excitation may be
localized on a defect. There are no other localized excitations
or they have a very low binding energy and are unstable. The dependence
of the density of localized states on the energy was chosen in the
exponential form, namely $\rho (E) = \alpha N_l \exp( \alpha E)$,
where $N_l $ is the density of the defect centers, $E$ is the
depth of the trap level. The exciton states (free and localized)
are distributed onto levels after the creation of electrons and holes
due to an external irradiation and their subsequent recombination and
relaxation. Since the time of relaxation is much less than
the exciton lifetime, the distribution of excitation between free
and localized states corresponds to  the thermodynamical
equilibrium state. In the considered model, we should obtain a
distribution of electron-hole pairs,  whose population on a
single level may be changed from zero to infinity for $E>0$ (for
 free exciton states) and from zero to one for $E<0$ (for
localized states). Formally, in the considered system,  free
excitons have Bose-Einstein statistics while localized excitations
obey the Fermi-Dirac statistics. At a small exciton density,
Bose-Einstein and Boltsmann statistics give similar results for free excitons,
  but the application of Fermi-Dirac statistics for localized states on a  single level for one trap is important.
   The equation for energy distribution
may be found from the minima of a large canonical distribution
\begin{equation}
\label{eqa} w(n_k,n_i)=\exp\left(\frac{\Omega+N\mu-E}{\kappa
T}\right),
\end{equation}
where $N=\Sigma_in_i+\Sigma_kn_k$,
$E=\Sigma_in_iE_i+\Sigma_{k,l}n_kE_k$, $n_i=0,1$,
$n_k=0,1,\ldots,\infty$, $k$ is the wave vector of the exciton, $l$
designates the singular levels. Parameter $\mu$ is the exciton chemical
potential.

\begin{figure}[!b]
\centerline{
\includegraphics[width=8.6cm]{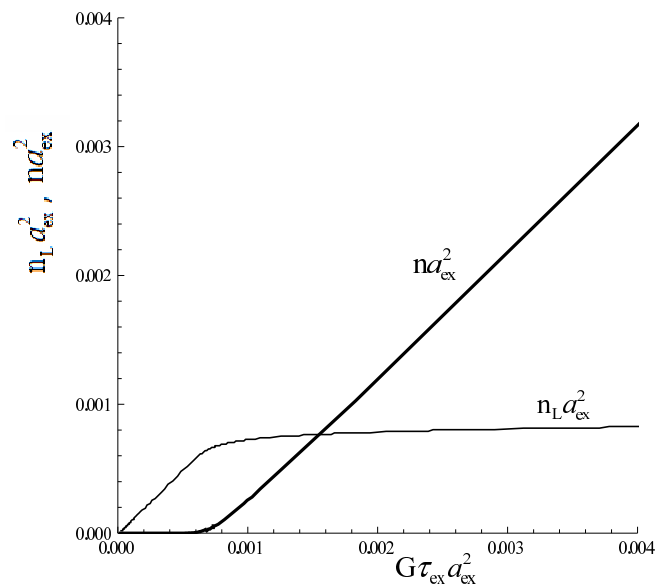}
}
\caption{
The dependence of the density of free (thick line) and trapped
(thin line) excitons on the pumping. The parameters of the system:
$T=2K,  N_l=0.001/a_\textrm{ex}^2,
 \alpha=300 (\textmd{eV})^{-1}$.}
 \label{fig3}
\end{figure}

The distribution of excitons over free and localized levels is
determined from the minimum of the functional (\ref{eqa}). As a
result, we obtain the following conditions for the mean values of
the  free exciton density $n$ and the density of localized
states $n_L$
\begin{align} \label{eqb} n_\textrm{ex}&=\frac{g \nu}{4\pi
E_\textrm{ex}a_\textrm{ex}^2}\int_0^\infty\frac{\rd E}{\exp\left(\frac{E-\mu}{\kappa
T}\right)-1} \,,\\
\label{eqc} n_L&=\alpha N_l
\int_{-\infty}^0\frac{\exp\left(\alpha E\right) \rd E}{\exp\left(\frac{E-\mu}{\kappa
T}\right)+1} \,,
\end{align}
where $a_\textrm{ex}=(\hbar^2\varepsilon)/(\mu_\textrm{ex} e^2)$ and
$E_\textrm{ex}=(\mu_\textrm{ex} e^4)/(2\varepsilon^2 \hbar^2)$ are the radius
and the energy   of the exciton in the ground state in the bulk
material,  $g=4$, $\mu_\textrm{ex}$ is the reduced mass of the exciton,
$\nu$ is the ratio of the reduced and the total mass of the
exciton. The chemical potential $\mu$ is determined from the condition
\begin{equation} \label{eq6}
n_L + n = G\tau_\textrm{ex}\,,
\end{equation}
\noindent where  $G\tau_\textrm{ex}$
 is the whole number of excitation (free and localized) per unit surface.

The dependence of distribution of free and localized
excitons on the pumping is presented in figure~\ref{fig3} as a function of the whole
number of excitation presented in  units of $1/a_\textrm{ex}^2$.
Let the exciton radius be equal to 10~{nm}. Then, the
concentration of the traps and the width of the distribution of
trap levels, chosen under calculations of figure~\ref{fig3}, are of the order of
$10^9$~{cm}$^{-2}$  and 0.003~{eV}, correspondingly.

\begin{figure}[!t]
\centerline{
\includegraphics[width=8.9cm]{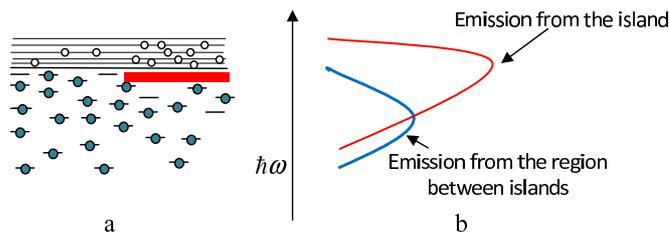}
}
\caption{
(Color online) The distribution of excitations in the traps and in the states of
the exciton band. The thick line in figure~\ref{fig4}~(a) corresponds to the energy
per a single exciton in the condensed phase. On the right [figure~\ref{fig4}~(b)],
the upper line describes the whole  emission from the island (the
emission of both the condensed phase and trapped excitons), the
low line describes the emission of the trapped excitons.
}
 \label{fig4}
 \end{figure}

As it is seen in figure~\ref{fig3}, the number of localized
excitations  at small pumping exceeds the number of free excitons
and the emission band is determined by the emission from the
traps. With an increase of  pumping, the occupation of the trap
levels becomes saturated. For the chosen parameters,  the concentration
of excitations under saturation is of the order of
$10^9$~{cm}$^{-2}$. The exciton density grows
simultaneously with the saturation of the localized levels. As
a result, the shortwave part of the emission band should
increase with an increased pumping. When the exciton density
 becomes larger, the collective exciton effects  begin to manifest themselves. The equations (\ref{eqb}), (\ref{eqc}) do not
 take into account the interactions between excitations, and special models and theories are needed
  to describe  the collective effects.
 The appearance of a narrow line was observed in \cite{Lar00} with an increased pumping on the shortwave
part of  the exciton emission band. Simultaneously, patterns
arise in the emission spectra. The narrow line appeared after the
localized states become occupied. According to \cite{Lar00}, this
line is explained by the exciton Bose-Einstain condensation.
According to our model \cite{Sug05,Sug07}, the appearance of the
islands corresponds to the condensed phase caused by the
attractive interaction between excitons.
The energy per a single exciton in the condensed phase
is less than the energy of free excitons (the thick line in
figure~\ref{fig4}), but the gain of the energy under condensation of indirect
excitons in AlGaAs system is less than the whole bandwidth,
which are formed by the localized and delocalized states. Thus, the
energy of photons emitted from the islands of a condensed phase is
higher than the energy of photons emitted by traps (see figure~\ref{fig4}).
The excitons cannot leave the condensed phase (the islands) and
move to the traps (to the states of lower energy) since the
levels of the traps are already occupied.  This  may be the reason of
the results obtained in \cite{Yan07}, where the maximum of the
frequency of emission from the islands is higher than the
maximum frequency from the regions between the islands \textit{in
spite of the attractive interaction between the excitons.}

The qualitative results coincide with the results obtained in
\cite{SugUJP} using another method from the solution of kinetic
equations for level distributions  at some simple approximation
for the probability transition between the levels. Similar
behavior of distribution of free and trapped excitons is observed for another energy dependence of the density of localized
states.

 The results may be used to explain the
intensity and temperature dependencies  of the exciton emission of
dipolar excitons in InGaAs coupled double quantum wells
\cite{Schin13}. The authors observed a growth of the shortwave
side of the  band with an increased pumping.

\section{Conclusion}

Hydrodynamic equations are analyzed for
excitons in a double quantum well. The equations take into account
the presence of pumping, the finite value of the exciton lifetime
and the possibility of a condensed phase formation in  a
phenomenological model. The equations describe the diffusion and
the ballistic movement of an exciton system. It is shown that
 the spatial nonuniform structures, observed  experimentally in double
wells on the basis of AlGaAs crystal, may be explained by
hydrodynamic equations in the diffusion approximation. The
effect of saturable localized states on spectral
distribution of the emission from condensed and gas phases is
obtained. The theory explains the features of experimental
dependencies of the emission spectra from the condensed and gas
phases.

\clearpage


\ukrainianpart

\title{Конденсація екситонів в квантових ямах. Гідродинаміка екситонів. Вплив локалізованих станів дефектів}
\author{В.Й. Сугаков}
\address{Інститут ядерних досліджень, просп. Науки, 47,
 03680 Київ, Україна}
%
%
%

\makeukrtitle

\begin{abstract}
\tolerance=3000%
{Проведено аналіз рівнянь гідродинаміки екситонів у квантовій
ямі. Рівняння враховують 1) можливість фазового переходу в
системі,  2) присутність зовнішньої накачки, 3) скінчений час життя
екситонів, 4) розсіяння екситонів на дефектах. Визначно
порогову накачку утворення періодичного розподілу екситонної
густини. Досліджується вплив локалізованих і вільних екситонів на
формування спектрів випромінювання.}
\keywords самоорганізація, квантові ями, екситони, фазовий перехід

\end{abstract}

\end{document}